\documentclass[twocolumn,showpacs,pra]{revtex4}
\usepackage{amssymb}
\usepackage{graphicx}
\begin{document}
\newcommand{\bra}[1]{\langle #1|}
\newcommand{\ket}[1]{|#1\rangle}
\newcommand{\braket}[2]{\langle #1|#2\rangle}
\newcommand{\sgn}{\mathrm{sgn}}
\title{Elementary Excitations of a Bose-Einstein Condensate in an Effective Magnetic Field}
\author{D. R. Murray}
\affiliation{Dept. of Physics, SUPA, University of Strathclyde, Glasgow G4 0NG, UK}
\author{P. \"Ohberg}
\affiliation{Dept. of Physics, SUPA, Heriot-Watt University, Edinburgh EH14 4AS, UK}
\author{Dami\'{a} Gomila}
\affiliation{Instituto de F\'{i}sica Interdisciplinar y Sistemas Complejos  (IFISC,CSIC-UIB), Campus Universitat Illes Balears, E-07122 Palma de Mallorca, Spain}
\author{Stephen M. Barnett}
\affiliation{Dept. of Physics, SUPA, University of Strathclyde, Glasgow G4 0NG, UK}

\begin{abstract}
We calculate the low energy elementary excitations of a Bose-Einstein Condensate in an effective magnetic field.  The field is created by the interplay between light beams carrying orbital angular momentum and the trapped atoms \cite{magfield05}.  We examine the role of the homogeneous magnetic field, familiar from studies of rotating condensates, and also investigate spectra for vector potentials with a more general radial dependence.  We discuss the instabilities which arise and how these may be manifested.
\end{abstract}
\maketitle

\section{Introduction}
Quantum degenerate gases are in many ways the ideal quantum many-body system.  In an experimental situation they afford an unprecedented level of control over the system parameters, such as the strength (and even sign) of the interaction between the atoms, the geometry of the external trap and the properties of the lattice in which the atoms are loaded.  It is no surprise therefore that Bose-Einstein Condensates (BECs) and degenerate Fermi gases are often used as a laboratory to study a host of phenomena from many different areas of physics.  This is especially true in condensed matter physics; for example, ultracold atoms in an optical lattice can be studied using the Hubbard model \cite{Jaksch98}.  Similarly, a system of trapped fermions tightly confined in one direction invites obvious comparisons with the 2D electron gas \cite{ohbergLandau}.

Without doubt, some of the most striking effects in solid state physics are observed when an external magnetic field is applied to a collection of charged particles.  Well known examples include the quantum Hall effects in 2D electron gases and the Meissner effect in Type II superconductors.  As the atoms forming quantum gases are electrically neutral, it is not obvious at a first glance how they might be used to study such effects.

The solution lies in the ability to create artificial magnetic fields.  For example, rotating the system and studying it in the rotating frame is analogous to studying charged particles in a homogeneous magnetic field \cite{Critrot02,rapidLLL,fastrot}.  Alternatively, lasers can be used to alter the state-dependent tunneling amplitudes of atoms in an optical lattice to simulate an effective magnetic flux \cite{zollerBeff,mueller}.

A recent proposal involves the adiabatic motion of lambda-type three level atoms interacting with laser-fields which create a non-degenerate dark state, that is an eigenstate of the atom-laser interaction.  It has been shown that if the atoms interact with a pair of laser beams possessing a relative orbital angular momentum \cite{OAMref,OAMbook}, then an effective vector potential appears in the effective equation for the atomic wavefunction \cite{ferfield,magfield05}.  The corresponding effective magnetic field created is entirely dependent on the form of the incident light, so that by appropriately choosing the light's phase and intensity  we can control both the strength and  shape (homogeneous or inhomogeneous) of the effective magnetic field.  The inherent flexibility of the system allows for wide-ranging studies into the magnetic properties of both degenerate Bose and Fermi gases, and could provide insight into gauge theories in general.

It is therefore pertinent to gain an understanding of how the fundamental properties of the gas may be modified in the presence of artificial magnetic fields.  A complete analysis must include the excitations, which determine the dynamical behaviour of the system under weak perturbation and are crucial in determining its superfluid properties.  Of particular interest are the lowest energy (or elementary) excitations, which are collective in nature.  In this paper we calculate the spectra  for a trapped 2D BEC in both homogenous and non-homogeneous magnetic fields, which are created as described in \cite{magfield05}.  Two-dimensional quantum gases have recently attracted a considerable interest in connection with the Kosterlitz-Thouless transition \cite{DalibardKT} and the quantum Hall effect in clouds of ultracold atoms (see \cite{ohbergLandau} and references therein).

The paper is organised as follows: in section II a brief description of the model is given and in section III we outline how the excitations are calculated.  As the interaction between the light and atoms introduces two effective potentials - a vector potential and also a scalar potential -  and both have a significant role to play in the dynamics and excitations,  we present our results in two parts.  In  section IV we assume the external trap has been chosen to counteract the effect of the additional effective trap so that the potential felt by the atoms is completely harmonic.  This allows us to isolate the role of the magnetic field alone on the excitations.  Then in section V we include the full effective trapping potential terms and study the excitations numerically. Finally in section VI we discuss and summarise the main results.

\section{The Model}

\begin{figure}[tbp]
\center{
\includegraphics[width=7cm]{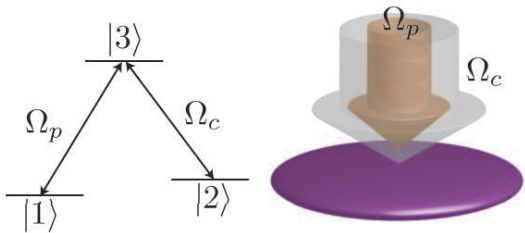}
\caption {Method for creating effective magnetic fields in degenerate atomic gases.  On the left is the level scheme for the $\Lambda$-type atoms interacting with the resonant probe beam $\Omega_{p}$ and control beam $\Omega_{c}$.  On the right is a schematic representation of the experimental setup with the two light beams incident on the cloud of atoms.  The effective magnetic field is generated if there is relative angular momentum between the beams.  This will occur, for example, if the probe field is of the form $\Omega_{p} \sim \rm{e}^{i\ell\phi}$, where each probe photon carries an orbital angular momentum $\hbar\ell$ along the propagation axis $z$, and $\Omega_{c}$ is independent of the azimuthal angle. }
\label{figlevels}}
\end{figure}

We consider a system of three-level atoms charecterized by two hyperfine ground levels $\ket{1}$ and $\ket{2}$ and an electronically excited level $\ket{3}$ interacting with two copropagating resonant laser beams in an EIT configuration (figure 1).  The probe beam, which has coupling strength $\Omega_{p}$ and is allowed to have angular momentum $\ell \hbar$ per photon along the z-axis, drives the transition $\ket{1} \rightarrow \ket{2}$, whilst the control beam has coupling strength $\Omega_{c}$ and is concerned with the transition $\ket{2} \rightarrow \ket{3}$.  These absorption paths destructively interfere to suppress transitions to level $\ket{3}$, driving the atoms to the dark state superposition of levels $\ket{1}$ and $\ket{2}$: $\ket{D}=\frac{\ket{1}-\zeta \ket{2}}{\sqrt{1+|\zeta|^2}}$, where $\zeta=\frac{\Omega_{p}}{\Omega_{c}}=\left |\frac{\Omega_{p}}{\Omega_{c}}\right | \mathrm{e}^{iS}$ and $S$ is the relative phase between the probe and control beam.  If the atoms in the dark state form a BEC, then the coupling between the light and the atoms introduces an effective vector potential into the mean-field equation for the atomic wavefunction $\Psi$ \cite{magfield05}:

\begin{equation}
i\hbar\frac{\partial\Psi}{\partial{t}}=\frac{1}{2M}(i\hbar\nabla+\mathbf{A})^{2}\Psi+V(\mathbf{r})\Psi+g \left| \Psi\right| ^{2}\Psi,
\label{TDGP1}
\end{equation}
where
\begin{equation}
\mathbf{A}=-\hbar \frac{|\zeta|^2}{1+|\zeta|^2}\nabla S
\label{A1}
\end{equation}
and
\begin{equation}
V=V_{ext}+\frac{\hbar^2}{2M}\frac{|\zeta|^2(\nabla S)^2+(\nabla |\zeta|)^2}{\left(1+|\zeta|^2\right)^2}
\label{V1}
\end{equation}
are, respectively, the effective vector and trapping potentials.  The external trapping potential for the dark state atoms is
\begin{equation}
V_{ext}=\frac{V_{1}+|\zeta|^2V_{2}}{\left(1+|\zeta|^2\right)^2}
\label{Vdark}
\end{equation}
where $V_{j}$ is the trapping potential for the atoms in hyperfine state $j$ ($j=1,2$).  The atomic interactions, which involve collisions between atoms both of the same and of different electronic state, can be described by a single parameter \cite{magfield05}
\begin{equation}
g=\frac{g_{11}+2g_{12}|\zeta|^2+g_{22}|\zeta|^4}{\left(1+|\zeta|^2\right)^2}.
\label{g}
\end{equation}
Here, $g_{ij}=\frac{4\pi\hbar^2a_{ij}}{M}$, where $a_{ij}$ is the $s$-wave scattering length between atoms in the levels $i$ and $j$ ($i,j=1,2$): $a_{jj}$ is the scattering length of atoms in the same electronic state and $a_{12}=a_{21}$ corresponds to collisions between atoms in different electronic states.  Note that in general the intractions can depend on position since $\zeta$ is position dependent.  However, if $\zeta$ is small or alternatively if the inter and intra-spieces scattering lengths are equal then $g$ is approximately constant throughout the condensate.  We shall assume this to be the case for the remainder of this paper.

For the purposes of our analysis it is convenient rewrite Eq. (\ref{TDGP1}) in the form
\begin{equation}
i\hbar\frac{\partial\Psi}{\partial{t}}=\left(-\frac{\hbar^2}{2M}\nabla^2+\tilde{V}+g \left| \Psi\right| ^{2} +\frac{i\hbar}{M}\mathbf{A}\cdot\nabla\right)\Psi,
\label{GProt}
\end{equation}
where $\tilde{V}(r)=V+\frac{|A|^2}{2M}$.  For the examples considered $\nabla \cdot \mathbf{A} =0$ so this equation is equivalent to (\ref{TDGP1}).

As can clearly be seen from Eq. (\ref{A1}), a non-vanishing effective magnetic field is created if there is relative  orbital angular momentum between the two light beams, such that the phase of the  dimensionless ratio $\zeta$ is given by $S=\ell \phi$, where $\phi$ is the azimuthal angle.  The shape of the effective vector potential is controled by the intensity ratio of the probe and control beams.  We choose $|\zeta|^2=\left|\frac{\Omega_{p}}{\Omega_{c}}\right|^2=\alpha_{0}\left(\frac{r}{R}\right)^{\nu+1}$, where the dimensionless parameter $\alpha_{0}$ is the ratio of probe to control beam at a characteristic radius $r=R$ which is chosen to be larger than the radius of the trapped cloud.  The exact forms of the resultant effective vector and trapping potential are then

\begin{equation}
\mathbf{A}=-\frac{\hbar \ell}{R}\frac{\alpha_{0}\left(\frac{r}{R}\right)^{\nu}}{1+\alpha_{0}\left(\frac{r}{R}\right)^{\nu+1}} \mathrm{e}_{\phi},
\label{exactA}
\end{equation}

and
\begin{eqnarray}
\tilde{V}=V_{ext}+\frac{\hbar^2}{2MR^2}\frac{[\ell^2+\frac{1}{4}(\nu+1)^2]\alpha_{0}\left(\frac{r}{R}\right)^{\nu-1}}{\left(1+\alpha_{0}\left(\frac{r}{R}\right)^{\nu+1}\right)^2} \nonumber \\
+\frac{\hbar^2}{2MR^2}\frac{\ell^2\alpha_{0}^2\left(\frac{r}{R}\right)^{2\nu}}{\left(1+\alpha_{0}\left(\frac{r}{R}\right)^{\nu+1}\right)^2}.
\label{exactV}
\end{eqnarray}

In order for the adiabatic dynamics to hold, so that the atoms remain in the dark state for a typical BEC lifetime, requires, typically, the ratio $|\zeta|^2 \lesssim 1$ and hence $\alpha_{0} \lesssim 1$  \cite{magfield05}.  If $\alpha_{0} << 1$ we can safely replace Eqs (\ref{exactA}) and (\ref{exactV}) by the approximate potentials

\begin{equation}
\mathbf{A}=-\frac{\hbar \ell}{R}\alpha_{0}\left(\frac{r}{R}\right)^{\nu} \mathrm{e}_{\phi},
\label{approxA}
\end{equation}

\begin{eqnarray}
\tilde{V}=V_{ext}+\frac{\hbar^2}{2MR^2}[\ell^2+\frac{1}{4}(\nu+1)^2]\alpha_{0}\left(\frac{r}{R}\right)^{\nu-1} \nonumber \\
-\frac{\hbar^2}{2MR^2}\ell^2\alpha_{0}^2\left(\frac{r}{R}\right)^{2\nu},
\label{approxV}
\end{eqnarray}
and the effective interaction strength (eq. \ref{g}) is constant throughout the cloud: $g \approx g_{11}$.  We obtain Eqs. (\ref{approxA}) and (\ref{approxV}) simply by expanding eqs (\ref{exactA}) and (\ref{exactV}) and neglecting all higher order terms in $\alpha_{0}$, except of course the 2nd term on the right hand side of (\ref{approxV}), because this is multiplied by an $\ell$ which can in principle be arbitrarily large.

The freedom to choose the form of the spatially-varying effective magnetic field relies on the ability to shape the intensities and phases of the incident laser beams.  Recent advances in light beam shaping, using for instance spatial light modulators, makes it possible to consider truly exotic light beams \cite{JohannesNature}.  The tightly confined or two dimensional gas offers in this respect a clear simplification: shaping a light beam in a plane is much less restrictive than in three dimensions, although 3D light shaping is certainly possible.

\section{Calculating the Excitations}

We calculate the excitations of this system by considering small time-dependent variations of the condensate wavefunction around the ground state $\Psi({\bf r})$, writing the wavefunction as
\begin{equation}
\Psi(\mathbf{r},t)=[\Psi(\mathbf{r})+u(\mathbf{r})\mathrm{e}^{-i\omega t}-v^{\ast}(\mathbf{r})\mathrm{e}^{i\omega t}]\mathrm{e}^{-i{\mu}t/\hbar}.
\label{fluct}
\end{equation}
Substituting this into eq. (\ref{GProt}) and keeping only linear terms in $u$ and $v$ we obtain two coupled equations analogous to the Bogoliubov - de Gennes equations \cite{Dalreview}
\begin{eqnarray}
&\left(\mathcal{L}_{0}-\frac{\hbar m}{M}\frac{A}{r}\right)u+g \left| \Psi\right| ^{2}(2u-v)=(\mu+E)u
\label{BDGu}\\
&\left(\mathcal{L}_{0}+\frac{\hbar m}{M}\frac{A}{r}\right)v+g \left| \Psi\right| ^{2}(2v-u)=(\mu-E)v,
\label{BDGv}
\end{eqnarray}
where $\mathcal{L}_0 = \frac{-\hbar^2}{2M}\nabla^{2}+\tilde{V}$and the (non-vortex) ground state satisfies
\begin{equation}
\left(\mathcal{L}_{0}+g \left| \Psi\right| ^{2}\right)\Psi=\mu\Psi.
\label{ground}
\end{equation}
These are the equations we solve to calculate the eigenfrequencies $\omega$ and eigenenergies $E=\hbar\omega$ of the excitation modes $u$ and $v$, which satisfy the normalization condition

\begin{equation}
\int d\mathbf{r}\left(u_{a}u_{b}^{\ast}-v_{a}v_{b}^{\ast}\right)=\delta_{ab}.
\label{normalization}
\end{equation}

In deriving Eqs. (\ref{BDGu})-(\ref{ground}) we have allowed the excitations to have angular momentum $m\hbar$ (where $m=0,\pm 1, \pm 2  ...$) by transforming $u$ and $v$ such that $u\rightarrow u e^{im\phi}$ and $v\rightarrow v e^{im\phi}$.  We assume that the ground state $\psi=\psi(r)$ only; this is equivalent to saying that $\psi$ does not correspond to a vortical state.  Note that a more riogorous approach for deriving the excitation equations would be to diagonalize the many body Hamiltonian in the Bogoliubov approximation, expressing the fluctuation operator in terms of quasiparticle operators \cite{ohbergetal97}.  The resultant equations which must be solved are identical to (\ref{BDGu})-(\ref{BDGv}).

The excitation spectrum of eqs (\ref{BDGu})-(\ref{BDGv}) can be calculated in the spirit of the Thomas-Fermi approximation where the effective trap potential (which we assume to have a harmonic ($\sim r^2$) component) and the repulsive mean-field interactions provide the dominant energy scales.  In this regime
$\mu\approx n_{max}g>>\hbar\omega_{t}$, where $n_{max}$ is the maximum condensate density.  We rescale the radial coordinates as $\rho=\frac{r}{L}$ and $\tilde{R}=\frac{R}{L}$ where $L=\left(\frac{2\mu}{m\omega_{t}^2}\right)^\frac{1}{2}$ is the characteristic length-scale of the harmonically trapped condensate with trap frequency $\omega_{t}$.  Introducing the dimensionless eigenenergies $\epsilon_{\nu}=\frac{E_{\nu}}{\hbar\omega_{t}}$ and the dimensionless density $\bar{n}=\frac{|\Psi|^2}{n_{max}}$, we can re-write our equations as

\begin{eqnarray}
&\mathcal{L}_{0_{d}}u+(2u-v)\bar{n}=\left[1+2\gamma(\epsilon+\gamma \frac{m A_{d}}{\rho})\right]u
\label{LBDGudim}\\
&\mathcal{L}_{0_{d}}v+(2v-u)\bar{n}=\left[1-2\gamma(\epsilon+\gamma \frac{m A_{d}}{\rho})\right]v
\label{LBDGvdim}\\
&\mathcal{L}_{0_{d}}\psi+\bar{n}\psi=\psi,
\label{Lgrounddim}
\end{eqnarray}
with $\mathcal{L}_{0_{d}}=-\gamma^2\nabla^{2}
+\tilde{V}_{d}$ where, in the Thomas-Fermi regime, the parameter $\gamma=\frac{\hbar\omega_{t}}{2\mu}<<1$ will allow for further simplification of Eqs. (\ref{LBDGudim})-(\ref{Lgrounddim}).  The subscript $d$ in $\tilde{V}_{d}$ and $A_{d}$ denotes that dimensionless units are being used.

Equations (\ref{LBDGudim})-(\ref{Lgrounddim}) are reduced to two fourth order differential equations after introducing the functions $f_{\pm}=u \pm v$ and substituting $\bar{n}$ from Eq. (\ref{Lgrounddim}):
\begin{eqnarray}
\frac{1-\tilde{V}_{d}}{\epsilon+\frac{\gamma{m}
A_{d}}{\rho}}\left(-\nabla^2f_{+}+f_{+}\frac
{\nabla^2\Psi}{\Psi}\right)\nonumber \\
+\frac{\gamma^2}{2}\left[-\nabla^2\left(\frac{1}{\epsilon+\frac{\gamma {m}A_{d}}{\rho}}
\right)\left(-\nabla^2f_{+}+f_{+}\frac{\nabla^2\Psi}
{\Psi}\right)\right] \nonumber \\
+\frac{\gamma^2}{2}\left[\frac{3}{\epsilon+
\frac{\gamma{m}A_{d}}{\rho}}\left(-
\nabla^2f_{+}+f_{+}\frac{{\nabla^2}\Psi}{\Psi}\right)\frac{{\nabla^2}\Psi}{\Psi}\right]\nonumber \\
=2(\epsilon+\frac{\gamma{m}A_{d}}{\rho})f_{+},\label{f+}
\end{eqnarray}
and
\begin{eqnarray}
 -\nabla^2\left(\frac{1-\tilde{V}_{d}}{\epsilon+\frac{\gamma{m}A_{d}}{\rho}}\right)f_{-}+\frac{1-\tilde{V}_{d}}{\epsilon+\frac{\gamma{m}A_{d}}{\rho}}f_{-}\frac
{\nabla^2\Psi}{\Psi}\nonumber \\
 -\frac{\gamma^2}{2}\left[\nabla^2\left(\frac{1-\rho^2}{\epsilon+\frac{\gamma{m}A_{d}}{\rho}}\right)\left(-\nabla^2f_{-}+3f_{-}\frac
{\nabla^2\Psi}{\Psi}\right)
\right] \nonumber \\
+\frac{\gamma^2}{2}\left[\nabla^2\left(\frac{1-\rho^2}{\epsilon+\frac{\gamma{m}A_{d}}{\rho}}\right)\left(-\nabla^2f_{-}+3f_{-}\frac
{\nabla^2\Psi}{\Psi}\right)\right]\nonumber \\
=2(\epsilon+\frac{\gamma{m}A_{d}}{\rho})f_{-},
\label{f-}
\end{eqnarray}

In the Thomas-Fermi approximation, we neglect the kinetic energy term $\gamma^2\nabla^{2}\psi$ in Eq. (\ref{Lgrounddim}) to obtain the wavefunction
\begin{equation}
\Psi_{TF}=\sqrt{n_{max}(1-\tilde{V}_{d})},\;\;\;\;\;\Psi_{TF}\geq{0},
\label{psiTF}
\end{equation}
which can readily be substituted into Eqs. (\ref{f+}) and (\ref{f-}).  When considering the low energy excitations, with wavefunctions that vary over a scale comparable with the size of the condensate, we must for consistency also neglect terms proportional to $\gamma^2$ in (\ref{f+}) and (\ref{f-}).  Then, applying the ansatz $f_{\pm}=C_{\pm}\left(\epsilon+\frac{\gamma{m}A_{d}}{\rho}\right)^{\frac{1}{2}}\left(\frac{1-\tilde{V}_{d}}{\epsilon+\frac{\gamma{m}A_{d}}{\rho}}\right)^{\pm\frac{1}{2}}Q(\rho,\phi)$, we derive in the Thomas-Fermi regime the equation for the excitations of a 2D condensate with effective magnetic field:
\begin{eqnarray}
(1-\tilde{V}_{d})\nabla^{2}Q-(1-\tilde{V}_{d})\frac{\nabla^2 \Psi_{TF}}{\Psi_{TF}}Q \nonumber  \\ =2\left(\epsilon+\frac{\gamma{m} A_{d}}{\rho}\right)^{2}Q,
\label{exc2}
\end{eqnarray}
The relation between the normalization constants $C_{+}$ and $C_{-}$ can be obtained from Eqs. (\ref{LBDGudim}), (\ref{LBDGvdim}) and (\ref{exc2}):

\begin{equation}
C_{-}=\gamma C_{+}.
\label{normtrap}
\end{equation}

\section{Harmonic Effective Trapping Potential}

\subsection{Homogeneous Magnetic Field}

By a judicious choice of external trapping potential, a purely harmonic effective trap can be obtained, such that $\tilde{V}=\frac{1}{2}M\omega_{t} r^2$ \cite{magfield05}.
If the ratio of control to probe intensity in the transversal plane is of the form $|\zeta^{2}|\sim r^{2}$, then the exponent $\nu$ in eq. (\ref{approxA}) is 1. The effective vector potential is then
\begin{equation}
\mathbf{A}=-\frac{\hbar\ell\alpha_{0}}{R^2}r\hat{e}_{\phi},
\label{UnB}
\end{equation}
corresponding to a uniform magnetic field in the z-direction i.e. $\mathbf{B}=\nabla\times\mathbf{A}=-\frac{2\hbar\ell\alpha_{0}}{R^2}\hat{e}_{z}$.  In this case, the solution of (\ref{exc2}) is of the form $W=x^{\frac{m}{2}}P(x)\mathrm{e}^{im\phi}$, where $x=\rho^2$ and the radial function $P(x)$ is the solution of the hypergeometric equation

\begin{eqnarray}
x(1-x)\frac{d^2P}{dx^2}+\left((m+1)-(m+2)x\right)\frac{dP}{dx}\nonumber \\ +\left(\frac{1}{2}{\epsilon}^{2}-\gamma\alpha_{0}\ell{m}\epsilon-\frac{m}{2}\right)P(x)=0.
\label{hypA1}
\end{eqnarray}


\begin{figure}[tbp]
\center{
\includegraphics[width=7cm]{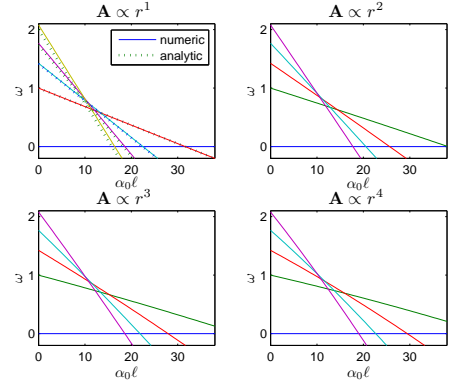}
\caption {Numerical results for surface mode (n=0) excitation frequency as a function of the effective magnetic field strength $\alpha_{0}\ell$ for different vector potentials.  The top left panel corresponds to the homogeneous magnetic field.  The lines correspond to the five lowest energy surface excitations, where the m=0 case (Goldstone mode) is represented by the lowest frequency mode at $\alpha_{0}\ell=0$, and the $m=4$ the highest.  Each mode exhibits an energetic instability beyond a critical $\alpha_{0}\ell$. The mode frequencies are purely real, so the system is dynamically stable when the vector and trapping potentials are radially symmetric.  A full description of the numerical method is given in section V.}
\label{fig-1}}
\end{figure}

For a physically well-behaved solution, we require $P(x)$ to be convergent at $x=0$ and converge as $x \rightarrow {1}$.  This yields the spectrum

\begin{equation}
E_{n,m}=\hbar\omega_{t}\left(2n^2+2n|m|+2n+|m|\right)^{\frac{1}{2}}+\frac{\hbar^2}{MR^2}\alpha_{0}\ell m
\label{spectrap1}
\end{equation}
where $n$ is the number of radial nodes and $m$ the angular momentum quantum number.  This spectrum is of the same form as that found by Ho and Ma \cite{HoMa} for the two-dimensional cloud, except for the shift term proportional to $\alpha_{0}\ell{m}$ which arises due to the effective magnetic field.  The uniform effective magnetic field is thus shown to induce a Zeeman-like shift on the energy levels, decreasing the excitation energy when $\ell$ and $m$ are opposite in sign and increasing it when they are the same.  The actual mode observed should be interpreted as a superposition of the + and - $m$ modes, as the effective magnetic field induces a rotation in modes with $m \neq 0$.  For example, if we excite the $m=\pm 1$ dipole mode (which corresponds to a sloshing motion along one axis) and then switch on the effective magnetic field the mode will start to precess at the effective cyclotron frequency  due to the additional force propotional to $\mathbf{v} \times \mathbf{B}$, where $\mathbf{v}$ is the velocity.

Let us also highlight that as a result of our choice of external trap, the spectrum of Eq. (\ref{spectrap1}) matches that of a rotating condensate when studied in the rotating frame, with rotation frequency given by $\Omega=\frac{\hbar \alpha_{0} \ell}{MR^2}$.  The stability properties of this system have been studied extensively (see e.g. \cite{SinhaCastin,Recati2001,UedaNonlin}).

The solutions of Eq. (\ref{hypA1}) are the Jacobi polynomials $P_{n}^{(|m|,0)}(1-2x)$ and from Eqs. (\ref{normalization}) and (\ref{normtrap})  we obtain
\begin{eqnarray}
f_{\pm}=\left[\frac{4n+2|m|+2}{L^2}\right]^{\frac{1}{2}}\left[\frac{(1-\rho^2)}{\gamma(\epsilon_{nm}-\gamma\alpha_{0}\ell m)}\right]^{\pm\frac{1}{2}}\nonumber \\ \times \rho^mP_{n}^{(|m|,0)}(1-2\rho^2).
\label{eig-untrap}
\end{eqnarray}

\subsection{Vector Potential for which the Magnetic Field is Zero}

If instead the ratio of the control and probe beam intensities were constant ($\nu=1$) we would obtain the vector potential
\begin{equation}
\mathbf{A}=-\hbar\alpha_{0}\frac{\ell}{r}\hat{\mathrm{e}}_{\phi}
\label{nonunA}
\end{equation}
in which case $\mathbf{B}=\nabla\times\mathbf{A}=0$ so that the effective magnetic field is zero throughout the cloud (in the same way that a velocity field proportional to $1/r$ around a vortex still satisfies the irrotationality criterion for Bose-Einstein Condensation).  An approximate energy spectrum can be derived by treating the effective potential as a small perturbation in eq. (\ref{exc2}), noting that this treatment breaks down as $r \rightarrow 0$.  The effecive vector potential plays the role of an additional centrifugal potential.  The solution of (\ref{exc2}) is of the form $W=x^{\frac{1}{2}\sqrt{m^2+4\gamma\alpha_{0}\ell{m}}}P(x)\mathrm{e}^{im\phi}$, where $x=\rho^2$ and $P(x)$ is governed by a hypergeometric equation which admits a physical solution convergent at the origin and as $x\rightarrow{1}$ only if
\begin{equation}
\frac{1}{2}\epsilon^2-\gamma\alpha_{0}\ell{m}\epsilon-(n+\frac{1}{2})\sqrt{m^2+4\gamma\alpha_{0}\ell{m}\epsilon}=n^2+n,
\label{solveeps}
\end{equation}
from which we can obtain the excitation spectrum by solving for $\epsilon$.  In the perturbative regime where $\epsilon\gamma\alpha_{0}\ell<<\frac{|m|}{4}$ we find
\begin{equation}
E_{n,m}=\hbar\omega_{t}\left(\epsilon^{(0)}+\frac{\hbar\omega_{t}}{2\mu}\alpha_{0}\ell\left(m+(2n+1)\sgn(m)\right)\right),
\label{spec-nonuntrap}
\end{equation}
where $\epsilon^{(0)}=\left(2n^2+2n|m|+2n+|m|\right)^{\frac{1}{2}}$ gives the spectrum when there is no vector potential and $\sgn(m)=+1,-1,0$  for $m$ positive, negative or $0$ respectively.  This is a somewhat crude approximation but, nevertheless, it yields an insightful result: the effective vector potential has significance even if the corresponding effective magnetic field seen by the atoms is zero, reminiscent of the Aharanov-Bohm effect.  This significance is manifested in a shift in excitation energy levels for all modes with angular momentum ($m \neq 0$), and the magnitude of the shift now also depends on $n$, the number of radial nodes.  The origin of the dependence on $\sgn(m)$ may have a topological interpretation.  

\subsection{Inhomogeneous Magnetic Field}

We can also use a perturbative approach on eq. (\ref{exc2}) to calculate the first order energy shift due to vector potentials with $\nu \ge 1$, which correspond to inhomogeneous effective magnetic fields.  The eigenfunctions and eigenvalues of the unperturbed Hermitian Hamiltionian $\hat{H}^{(0)}=\left(1-\rho^2\right)\nabla^{2}-2\rho\frac{\partial}{\partial\rho}$ are those of the harmonically trapped BEC when no effective magnetic field is present: $Q^{(0)}=\rho^{m}P_{n}^{(|m|,0)}(1-2\rho^2)\mathrm{e}^{i m \phi}$ and $\epsilon^{(0)}=\left(2n^2+2n|m|+2n+|m|\right)^{\frac{1}{2}}$, where $\epsilon^{(0)}$ is degenerate with respect to the sign of $m$.  The first order energy corrections due to the perturbation $H'=\gamma m |\mathbf{A}| = \frac{\gamma m \alpha_{0} \ell}{\tilde{R}^2} \left(\frac{\rho}{\tilde{R}}\right)^{\nu-1}$ are the solutions of the secular equation

\[  \left| \begin{array}{ccc}
H'_{++}-\epsilon_{r}^{(1)} & H'_{+-}  \\
H'_{-+} & H'_{--}-\epsilon_{r}^{(1)} \end{array} \right|=0\]
with $H'_{\pm\pm}=\kappa_{r}\bra{Q_{\pm}^{(0)}} H' \ket{Q_{\pm}^{(0)}}$, where $+$ or $-$ denotes the sign of $m$, and the constant $\kappa_{r}$ can be determined using Eqs. (\ref{normalization}) and (\ref{normtrap}).  The off-diagonal terms vanish, allowing us to express the first order energy shift $\Delta E=E^{(1)}-E^{(0)}$ of the surface modes ($n=0$) as

\begin{figure}[tbp]
\center{
\includegraphics[width=7cm]{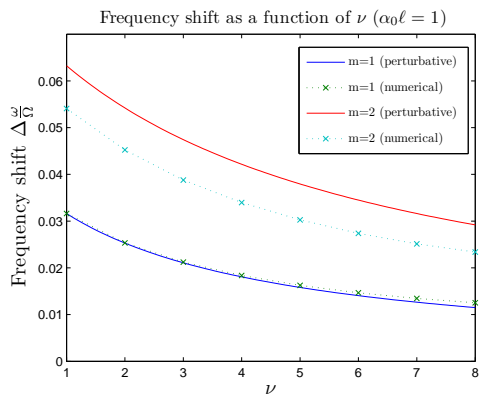}
\caption {Frequency shifts for the $m=1$ and $m=2$ modes due to the presence of the effective magnetic field as a function of $\nu$, the radial exponent in the vector potential (Eq. (\ref{approxA})).  The dotted lines correspond to results obtained solving Eqs. (\ref{LBDGudim})-(\ref{Lgrounddim}) numerically and the solid lines to the results of the perturbative calculation (Eq. (\ref{perturbshift}).  All frequencies are in units of the harmonic trap frequency $\omega_{t}$, and the parameters used were $\alpha_{0}\ell=-1$ and $\gamma^2=0.001$, which describes a condensate well in the Thomas-Fermi regime. }}
\label{fig-3}
\end{figure}

\begin{equation}
\Delta E=\frac{\hbar^2\alpha_{0}\ell m}{MR^2}\frac{2(|m|+1)}{\left(2|m|+\nu+1\right)},
\label{perturbshift}
\end{equation}
which is plotted in  figure 3.  The perturbative results agree well with our numerical calculations for the lowest energy excitations but breaks down as we increase $n$ and $m$ as would be expected since these excitations are no longer slowly varying in space.

The observed shift on the excitation energy levels is reduced as we increase the exponent term $\nu$ in the vector potential from 1.  This is not surprising as the radial coordinate $\rho$ is $\le 1$ and therefore for larger $\nu$ the effective magnetic field becomes more concentrated at the edge and permeates less to the centre of the cloud.  As a consequence, the critical $\alpha_{0}\ell$ for energetic instabilities to occur increases as the exponent $\nu$ is increased.  However, the perturbative treatment is valid only for $\frac{E^{(1)}}{\Delta E} << 1$ and so cannot be used to predict energetic instabilities.  These instabilities can be inferred from the numerical calculations of the spectra for $\nu=2,3,4$, presented in figure 2. A full description of the numerical method is given in the next section.
In contrast to the case of homogeneous effective magnetic fields we also note that the magnitude of the energy shift depends on the radial node index $n$.  The effective magnetic field does not, however, affect the purely compressional ($m=0$) modes with zero angular momentum.

\section{Full Effective Trapping Potential induced by the light}

We now present the results of numerical calculations used to determine the excitations when the only assumption we make of our external trap is that it is harmonic, and include the additional effective trapping potential due to the interaction with the light and atoms.  For realistic parameter values, with $\ell > 1$ and $\alpha_{0} < 1$, the effect of the trapping potential induced by light eclipses that of the effective magnetic field, with profound implications for the excitation spectra.

The exception is the case where $\nu=1$.  If $\alpha_{0}$ is small, then the vector potential approximated by Eq. (\ref{approxA}) corresponds to a homogeneous magnetic field.  The second term in Eq. (\ref{approxV}) represents a uniform shift in the chemical potential throughout the cloud - an effect we can ignore by setting the effective trap minimum to zero - while the third term shifts the harmonic trap frequency downwards.  The spectrum is then the same as Eq. (\ref{spectrap1}) except that we replace the trap potential $\omega_{t}$ such that
\begin{equation}
\omega_{t} \rightarrow \left(\omega_{t}^2-\left(\frac{\hbar\alpha_{0}\ell}{MR^2}\right)^2\right)^\frac{1}{2}.
\label{freqshift}
\end{equation}
\begin{figure}[tbp]
\center{
\includegraphics[width=7cm]{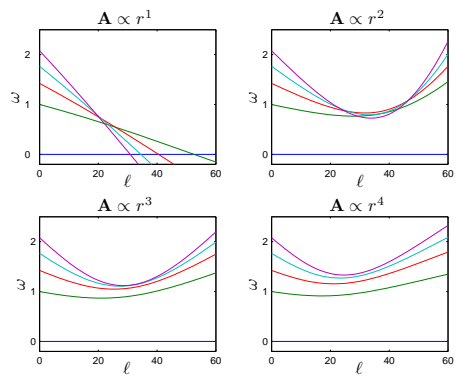}
\caption {Numerical results for surface mode ($n=0$) excitation frequency as a function of the effective magnetic field strength $\alpha_{0}\ell$ for different vector potentials with radial exponent $\nu$. The lines correspond to the five lowest energy surface excitations, where the $m=0$ case is represented by the lowest frequency mode at $\alpha_{0}\ell=0$, and the $m=4$ the highest.  Only the $\nu=1$ case exhibits energetic instabilities.  Here $\alpha_{0}=0.5$, $\gamma^2=0.001$ and $\tilde{R}=1$. The number of grid points used was 128 and the system size was L=2.56}}
\label{fig-4}
\end{figure}

The spectrum exhibits energetic instabilities above a critical angular momentum of the light $l$, as in section IV.  However, this is not the case for $\nu>1$.  In figure 4 we plot the dispersion curves obtained for $\nu=1,2,3,4$.  An analytical approach is no longer possible when  $\nu >1$, corresponding to an inhomogeneous effective magnetic field, and/or if the intensity ratio of the control and probe beam is such that the inequality $\alpha_{0} \ll 1$ is violated. In these cases we calculate the excitations solving Eqs. (\ref{LBDGudim})-(\ref{Lgrounddim}) numerically using the exact expressions for the
potentials (Eqs. (\ref{exactA}-\ref{exactV})).  First,  we solve the radial version of Eq. (\ref{Lgrounddim}) imposing first
derivatives equal to zero at $r=0$ and $r=L$, with $L$ the domain size. To do so we discretize the space and solve the set of coupled ordinary differential
equations using a Newton method in which the derivatives are computed in
Fourier space \cite{McSloyetal,Gomilaetal}. Solving Eqs. (\ref{LBDGudim}) and (\ref{LBDGvdim}) then reduces to finding the
eigenvalues of the discretized matrix associated to the linear problem
(\ref{LBDGudim}-\ref{LBDGvdim}).  For $\nu \ge 2$ the mode frequencies only decrease to some value above zero until an intermediate value of $\ell$, above which they increase monotonically with $\ell$.

\section{SUMMARY AND DISCUSSION OF MAIN RESULTS}

In the studies of BECs subject to an external rotation - equivalent to applying a homogeneous effective magnetic field - the presence of instabilities in the excitation spectra has been shown to play a crucial role in determining the evolution of the condensate, particularly with respect to the nucleation of vortices.  There are two types of instability which ought to be considered.  Dynamic instabilities are associated with a complex excitation energy and departure from the initial configuration due to interaction effects.  The energetic instability relates to the existence of excitations with negative energy, so that in the presence of dissipation the system can lower its energy by going into an `anomolous' mode, and is a prerequisite for dynamical instability \cite{lesHouches2}.  The dynamical instability has been credited as the primary mechanism for vortex nucleation in the rotating trap experiments \cite{SinhaCastin}.  It has also been argued that while the dynamic instability helps induce vortex nucleation, the actual penetration of vortices in to the bulk is a consequence of the energetic instability \cite{UedaNonlin}.

The excitation spectra studied in this paper, both analytically and numerically do not exhibit any dynamic instabilities in that we do not observe a critical effective magnetic field at which the excitation energy becomes complex.  Due to the increased complexity of adding an inhomogeneous effective magnetic field, however, we have restricted ourselves to consider trapping configurations which are radially symmetric.  For the homogeneous effective magnetic field dynamical instabilities are only observed with the introduction of an anisotropic trap \cite{SinhaCastin}, and this is likely to be the case for inhomogeneous magnetic fields as well.

Energetic instabilities, by contrast, occur readily.  For all the vector potentials considered in section IV the numerical results display energetic instabilities beyond a critical field strength proportional to $\alpha_{0} \ell$.  As we increase the radial exponent $\nu$ in the vector potential we in turn must significantly increase the probe beam angular momentum before the instability can be observed.  When the full effective trapping potential is included the $\nu=1$ case exhibits an energetic instability but in the vast majority of parameter space those for $\nu > 1$ do not.  Only in the region where the maximum intensity ratio $\alpha_{0}$ approaches 1 do we observe instabilities in the low-energy modes.  However, it is in this region the adiabatic dynamics most easily breaks down \cite{magfield05} and the model described in section II may become invalid.

Another important feature of the spectra calculated in sections IV and V is the accidental degeneracies, which occur where the mode frequencies intersect in figures 1 and 3.  These degeneracies are likely to be manifested through transfer of excitations from higher order modes to lower order modes at the point of degeneracy.  This kind of phenomenon could have implications for the condensate evolution depending on whether the effective magnetic field is switched on instantly or its strength is adiabatically ramped up to a final value.  For example, an anisotropically trapped condensate in a homogeneous effective magnetic field naturally undergoes $m=2$ quadrupole  oscillations, but the higher order modes exhibit dynamic instability \cite{SinhaCastin}, eventually leading to vortex nucleation.  To properly account for the mode-coupling we would need to move beyond Bogoliubov - de Gennes theory which does not account for  interactions between the degenerate modes \cite{ohbergetal97}.

With inclusion of the full effective trap induced by the interaction with the light we reach a scenario where there are no accidental degeneracies as $\nu$ is increased beyond 1.  The vector potential appearing in the Bogoliubov-de Gennes equations  is less important in determining the excitation frequencies in comparison with the effective trap.  A direct measure of the contribution of the vector potential is the splitting of the + and - $m$ modes. For the potentials with $\nu > 2$ studied in section V, we find that the splitting is typically $< 5\%$ of the actual mode frequency if no effective magnetic field were present.  This is in stark contrast to section IV where the splitting can easily exceed twice the mode frequency causing an energetic instability for realistic parameter values.

In this paper we have deliberately paid particular attention to the surface modes ($n=0$, $m \ne 0$), which are naturally excited by adding the effective magnetic field to an anisotropically trapped condensate \cite{DMvortex}.  We predict that for inhomogeneous effective magnetic fields vortex nucleation will be inhibited due to the dominance of the trapping potential induced by the light.  In general, it is the ratio of the cyclotron frequency to the effective trapping frequency which drives the dynamics of the system and both of these depend on $\alpha_{0} \ell$.  It is therefore often desirable to introduce a counter potential to act against the additional trapping terms due to the interaction with the light, as described in section IV.

\begin{acknowledgements}
This work was supported by the United Kingdom EPSRC. D.G. acknowledges financial support from MEC (Spain, Grant No. TEC2006-10009) and Govern Balear (Grant No. PROGECIB-5A).  Helpful discussions with G. Juzeli\ifmmode \bar{u}\else \={u}\fi{}nas and G -L. Oppo are gratefully acknowledged.
\end{acknowledgements}


\bibliographystyle{apsrev}

\end{document}